\documentclass[journal]{IEEEtran}
\IEEEoverridecommandlockouts
\usepackage{amsmath, amssymb, amsfonts, graphicx, algorithm, algorithmic, enumerate,color}
\usepackage{cite}
\usepackage{graphicx}
\usepackage{textcomp}
\usepackage{mathtools}
\usepackage{subcaption}

\usepackage{tikz}
\newcommand{\comment}[1]{}

\DeclareMathOperator{\E}{\mathbb{E}}
\def\BibTeX{{\rm B\kern-.05em{\sc i\kern-.025em b}\kern-.08em
    T\kern-.1667em\lower.7ex\hbox{E}\kern-.125emX}}

\begin{document}
\title{Joint Beamforming and Location Optimization for Secure Data Collection in Wireless Sensor Networks with UAV-Carried Intelligent Reflecting Surface}
%{Secure collaborative beamforming in distributed wireless sensor networks with UAV-Carried Intelligent Reflecting Surface: Joint Location and Beamforming Optimization}

\author{Christantus O. Nnamani, 
        Muhammad R. A. Khandaker,~\IEEEmembership{Senior Member,~IEEE,} and
        Mathini Sellathurai,~\IEEEmembership{Senior Member,~IEEE}

\thanks{Christantus Nnamani, Muhammad Khandaker and Mathini Sellathurai are with the School of Engineering and Physical Sciences, Heriot-Watt University, Edinburgh EH14 4AS, United Kingdom.}
%\thanks{R. T. Khan is with the Institute of Information Technology, Jahangirnagar University, Dhaka, Bangladesh.}

\thanks{This work was supported in part by the EPSRC Project EP/P009670/1, Petroleum Technology Development Fund and University of Nigeria Nsukka.}
\thanks{\texttt{Submitted Version. Copyright belongs to IEEE.}}
}

\maketitle

\begin{abstract} \label{abs}
This paper considers unmanned aerial vehicle (UAV)-carried intelligent reflecting surface (IRS) for secure data collection in wireless sensor networks. An eavesdropper (Eve) lurks within the vicinity of the main receiver (Bob) while several randomly placed sensor nodes beamform collaboratively to the UAV-carried IRS that reflects the signal to the main receiver (Bob). The design objective is to maximise the achievable secrecy rate in the noisy communication channel by jointly optimizing the collaborative beamforming weights of the sensor nodes, the trajectory of the UAV and the reflection coefficients of the IRS elements. By designing the IRS reflection coefficients with and without the knowledge of the eavesdropper's channel, we develop a non-iterative sub-optimal solution for the secrecy rate maximization problem. It has been shown analytically that the UAV flight time and the randomness in the distribution of the sensor nodes, obtained by varying the sensor distribution area, can greatly affect secrecy performance. In addition, the maximum allowable number of IRS elements as well as a bound on the attainable average secrecy rate of the IRS aided noisy communication channel have also been derived. Extensive simulation results demonstrate the superior performance of the proposed algorithms compared to the existing schemes.
\end{abstract}


\section{Introduction} \label{intro}
It is expected that the next generation networks will be equipped with more active and passive communication nodes to improve on coverage and capacity by accommodating the use of higher frequency spectrum (e.g. millimeter wave) and the deployment of massive multiple-input multiple-output (M-MIMO) systems \cite{jrnl_6g, wu_2020}. Nevertheless, these expectations foretell several logistic and environmental drawbacks ranging from increased communication cost in terms of energy, space and finance, to lower security guarantees. Hence, in recent times, secure, high capacity, energy efficient and cost effective communication systems have been the paramount facet in developing technologies for the next generation of wireless communication systems \cite{jrnl_6g, jrnl_secrecy}. These next generation technologies must ensure ultra-reliable connection to massive end user nodes amidst rapid time varying channels due to fast mobility and complex inter-connection \cite{jrnl_6g}. Presently, improvements to traditional technologies inclined to ultra connection and adapting to these growing requirements, only act as a conduit since they will definitely be overwhelmed in the advent of full deployment of the next generation systems of the internet of everything and artificial intelligence.

In view of this, several cost effective wireless communication techniques have been proposed in recent times such as lens MIMO, hybrid beamforming, unmanned aerial vehicle (UAV)/drone communications, advanced analog to digital converters (ADCs) etc. Recently, focus on wireless channel control has led to a shift in paradigm with the discovery of the intelligent reflecting surface (IRS) \cite{Basar_2019, Zhao_2019, conf_irs_izhuo}. The IRS is pitched to provide an interface between the traditional wireless base stations and the users. However, unlike the conventional active relays, the radio signal are only reflected by IRS and are inherently free from self-interference while traveling through a conditioned wireless communication channel \cite{wu_2020}. An overview of IRS with the technicalities of the physical implementation has been presented in \cite{wu_2020, Zhao_2019, Basar_2019}. Nevertheless, current literature explores several designs to harness the intelligence of the IRS for effective communication by optimally modelling the reflection coefficient and other related established parameters like beamforming weights where multiple sources/receivers are applicable.

For specific internet of things (IoT) applications, a joint optimization of the transmit beamforming and the reflection coefficients of the IRS system serving as relay between multiple antenna access point (AP) and a single antenna user was proposed in \cite{wu_2018}. This downlink communication technique was reversed and deployed to extend the coverage region of a base station by constructively reflecting its impinging signal to target locations either terrestrial \cite{Sixian_2020} or aerial \cite{Ma_2020}. However the proclivity of a multi antenna AP or base station in most IRS literature do not emphasis the deployment of IRS for noisy multi-sensor scenario that best depicts a typical IoT application module. These models also emphasize deployment of IRS to channel models with low noise component hence, do not represent practical application. Such practical IoT applications can range from pipeline monitoring to massive sensor interaction from self-driving cars. In this multi-sensor scenario, it is apparent that the beamforming weights will be determined collaboratively \cite{Zarifi_2010} since the ordered structure of the antenna element has been transformed to a random variant. %Other studies of IRS designs in wireless communication include energy efficiency maximization [15], secrecy rate maximization [17], joint active and passive beamforming design [18], and rate region characterization for IRS-aided interference channel [24], etc. 

With the proliferation of computing capabilities as well as wireless IoT applications, security of communications is becoming increasingly vulnerable to external threats \cite{jrnl_lwc_survey, jrnl_secrecy, jrnl_sec_sinr}. Accordingly, additional security measures including lightweight cryptography \cite{jrnl_lwc_survey} and physical layer security (PLS) have been proposed \cite{jrnl_secrecy, jrnl_sec_sinr, jrnl_sec_const, jrnl_const_swipt}, particularly for resource-constrained IoT devices. More recently, IRS-enabled programmable wireless channels have been proposed to make PLS even more practicable for the secrecy of communications \cite{Chen_2019, Akyildiz_2020}. However, the reflected signal from the IRS in radio spectrum cannot be said to be specular due to the dependence of its beamwidth on wavelength and IRS plate width \cite{Emil_2020}. Hence, the security of the communication cannot be perfectly guaranteed for radio signals especially if the channel state information of the eavesdropper is unknown. 
%Nevertheless, sequel to the extensive benefit of IRS deployment for capacity and rate improvement, it ensures positive secrecy rates even when the channel of the legitimate receiver and eavesdropper are highly correlated \cite{Cui_2019, Chen_2019}. 
However, due to the manipulation of the wireless channel to be favorable at desired location and adverse at another, the PLS of the IRS-aided communication is substantially better than modern conventional communication \cite{Cui_2019, Chen_2019}. By \comment{jointly }optimizing \comment{the beamforming weights and }the reflection coefficients in the presence of eavesdropper(s), the positive secrecy rate is obtained and can be substantially improved \cite{Cui_2019}. %\cite{Cui_2019} considered the presence of an eavesdropper in a communication from a MISO system  with fixed IRS system. 

One key feature of the secrecy IRS system is that the reflected signal can be possibly made to constructively sum at the legitimate receiver while destructively combined at the eavesdropper. However, it is relevant to note that the IRS placement is not necessarily optimal due to its fixed nature \cite{Lu_2020} and this can limit its performance in terms of PLS if the channels of the main receiver and the eavesdropper are correlated. Most of the existing works on IRS-aided communication focus on fixed terrestrial IRS deployment (on facades of buildings or indoor walls/ceilings) thereby promoting fixed IRS structures. This deployment does not provide adaptability to mobile users and do not guarantee that the IRS has been deployed at the optimal location\comment{\cite{Lu_2020}}, thereby undermining the full potential of IRS in terms of information rate and PLS. It is desirable to allow mobility of the IRS system to position at the best location possibly by mounting on a UAV or other aerial devices as recently studied in \cite{Lu_2020}. The choice of aerial mobility system also allows for the exploitation of aerial visibility for the IRS line-of-sight (LoS) application. One major drawback of aerial IRS is that malicious users can also establish an LoS link between the aerial IRS system and relying on the non-specular nature of the reflected radio signal, compromise the secrecy of the communication. The design of aerial IRS considering PLS has not been examined in the literature to the best of our knowledge. %Unfortunately, \cite{Lu_2020} did not consider the secrecy of the communication nor the source wireless sensor network thereby undermining the PLS of the aerial IRS system for typical IoT applications.

%Furthermore, as observed in [ ], this iterative process increases the complexity of the system. 

This motivates us to study in this paper the effect of UAV-carried IRS system on secure data transmission rate for multi-sensor IoT applications in a noisy environment. The objective is to maximise the achievable secrecy rate under total transmit power constraint by jointly optimizing the transmit beamforming weights, IRS reflection coefficients and the location of the IRS system aided by the mobility of the UAV. We assume that there is no direct link between the sensors and the base station (BS), therefore the communication link is established by deploying the UAV-carried IRS. The major contributions in this paper are listed below.
\begin{itemize}
\item[a)] We first design the IRS reflection coefficients with the knowledge of the main receiver's channel only and considering the worst-case scenario in determining the UAV trajectory. This enables us to avoid the conventional iterative procedure due to the inter-dependence of the transmit beamforming weights and the IRS reflection coefficients and thus reduce to a non-iterative solution approach. Extensive simulation results demonstrate that the non-iterative procedure yields comparable performance to the iterative procedure in terms of secrecy rate while reducing the computational complexity significantly. We then perform the same taking eavesdropper's channel information into consideration for the baseline scheme.

\item[b)] Furthermore, considering the noisy environment, we examine the influence of the number of IRS elements on the average secrecy rate and define a bound that links the number of IRS elements to the transmit signal wavelength, $\lambda$. 

\item[c)] Finally, we show that the time of flight of the UAV and the randomness in the distribution of the transmit sources (i.e. wireless sensors in this case), obtained by varying the sensor distribution area, generally improves the secrecy rate. 
\end{itemize}

To the best of our knowledge, no other existing work has investigated aerial/mobile IRS for secure communications with a view to define the maximum IRS elements for optimal secrecy rate in a noisy communication channel. Also, the proposed non-iterative solution for the IRS reflection coefficients and beamforming weights provides significant new insights to the problem. The contribution is therefore significantly novel.

%%%%%%%%%%%%%%%%%%%%%%%%%%%%%%%%%%%%%%%%%%%%%%%%%%%%%

\noindent{\bf Notations:} $\{\cdot\}^T$ and $\{\cdot\}^{\rm H}$ represent the transpose and Hermitian of vectors/matrices, respectively, while $\hat{\bf a}_{ij}\coloneqq \frac{{\bf a}_j-{\bf a}_i}{\|{\bf a}_j-{\bf a}_i\|}$ represents a normalized/unit vector along the direction of propagation from location $i$ to $j$. ${\rm diag}({\bf x})$ is a diagonal matrix with $\bf x$ as the main diagonal and ${\bf 1}$ is a column vector of $1$'s.

\begin{figure}[!ht]
\centering
\includegraphics[width=1.0\linewidth]{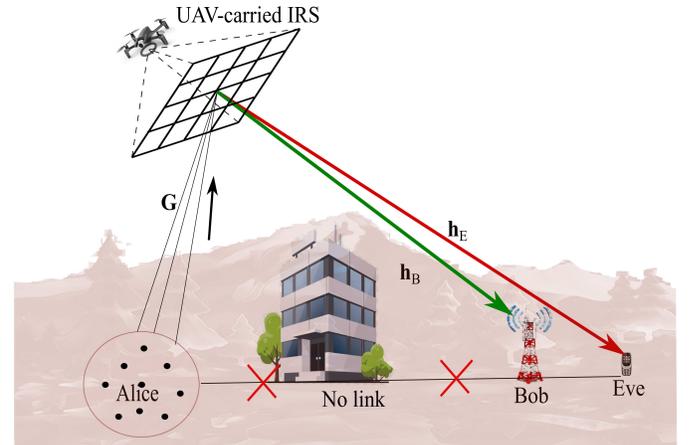}
\caption{Schematic of the UAV-IRS interaction with ground nodes}
\label{sys_m}
\end{figure}

\section{System Model and Problem Formulation} \label{sec2}
Let us consider that an IRS is being carried by a UAV tracing a path such that the reflected signals from transmitter (Alice) are received at a base station (Bob) which is physically incommunicado as shown in Fig. \ref{sys_m}. Since the radio signals from the IRS are not specularly reflected especially in noisy environment, (i.e. the reflection is not mirror-like \cite{Emil_2020}), an eavesdropper (Eve) lurking around Alice can receive an out-of-phase version of the reflected signals. This invariably compromises the secrecy of the communication between Alice and Bob, especially, if Eve has access to advanced signal reconstruction technologies.

Let us consider that Alice is a group of closely packed $M$ sensor nodes at the ground level within an area of radius $r$ that act collaboratively to beamform a unique symbol to Bob through the IRS. We denote the 3D location of Bob, Eve, and the center of Alice as $\boldsymbol{\Omega}_{\rm B}$, $\boldsymbol{\Omega}_{\rm E}$, $\boldsymbol{\Omega}_{\rm A}$, respectively. We note that Bob and Eve are in the far-field of the IRS system as required practically. Let the entire flight time ($T$) of the UAV be sampled at discrete time-stamps of $N$ equal time slots, with duration $\alpha = \frac{T}{N}$. Without loss of generality, we assume that the UAV flies at a constant altitude $H$ and a maximum speed of $Z$m/s for each $\alpha$ seconds, giving rise to a 3D trajectory represented as ${\bf Q}=\{{\bf q}[n]=[q_{\rm x}[n],q_{\rm y}[n], H]^T, n\in \{1,\dots,N\}\}$. We note that as $N \rightarrow \infty$, the UAV are seen as following a continuous trajectory satisfying time-sharing conditions, thereby, reflecting the signals continuously through its entire flight time \cite{uav_cooperative_jamming, obinna_2020}. For ease of computation, the IRS are placed on the UAV such that, for each $n\in\{1,\dots,N\}$, the location of the first IRS element, which we consider as the reference element, is the same as the location of the UAV $({\bf q}_{\rm R}[n]={\bf q}[n])$. Given that $\lambda$ is the carrier wavelength, all other adjacent elements of the IRS are separated by a fixed distance such that $d_{\rm x}< \frac{\lambda}{2} {\rm ~and~} d_{\rm y}< \frac{\lambda}{2}$ \cite{Lu_2020}. This implies that $d_{\rm x}= \frac{\lambda}{z_{\rm x}} {\rm ~and~} d_{\rm y}= \frac{\lambda}{z_{\rm y}},$ where $z_{\rm x}> 2 {\rm ~and~} z_{\rm y}> 2$. Hence we define the location of the $n$th IRS element such that ${\bf Q}_{\rm IRS}[n]=({\bf q}_{\rm Rx}[n]+(k_{\rm x}-1)d_{\rm x}, {\bf q}_{\rm Ry}[n]+(k_{\rm y}-1)d_{\rm y}, H),~ n\in \{1,\dots, N\},~ k_{\rm x}\in [1, \dots, K_{\rm x}], ~ k_{\rm y}\in [1, \dots, K_{\rm y}]$; where $K_{\rm x}$ and $K_{\rm y}$ define the number of IRS elements along the ${\rm x}$- and ${\rm y}$-directions of the grid, respectively. Since, the total number of IRS elements is given as $K=K_{\rm x}K_{\rm y}$, a compressed form of IRS element location can be written as ${\bf Q}_{\rm IRS}[n]=\{{\bf q}_k[n], \forall{n\in [1,\dots,N],~\&~ \forall{k}\in [1,\dots,K}]\}$ .

The UAV's continuous flight trajectory causes the reflected signals from the IRS to undergo Doppler shift and time variation at the receiver ground nodes (Bob and Eve). However, since all the IRS elements are fixed on the UAV and travelling at the same velocity, the Doppler shift due to their respective position will be uniform. Nevertheless, considering that the analysis to determine the UAV trajectory and secrecy of the communication is performed at instantaneous samples of $n$ where the UAV is assumed to be static, the effect of the Doppler shift vanishes as the UAV's velocity is zero at instantaneous $n$th sample point.

Following the convention as in \cite{Zarifi_2010}, we assume that the sensors in Alice collaboratively transmits a unique symbol $s(t)$ with ${\rm \E}\{|s(t)|^2\}=1$ giving rise to a passband signal of $ x(t)=s(t)\exp({j\omega_ct})$. The incident signal on the IRS elements from the $m$th sensor of Alice during an $n$th sampling period is given by
\begin{align}\label{rxu}
{\bf r}[n]={\bf g}_m[n]{w}_{m}[n]x(t)[n],% + {\bf n}_m[n], 
\end{align}
where ${\bf g}_m=[g_{m1},g_{m2},\dots,g_{mK}]^T$ %and ${\bf n}_m~\sim~ \mathcal{C}^{K\times 1}$ 
denote the IRS to the $m$th sensor complex channel vector %and the independent and identically distributed (i.i.d.) white Gaussian noise vector due to the line-of-sight (LoS) link between the IRS elements and the $m$th sensor, respectively, 
while $w_m$ represents the beamforming weight of the $m$th antenna element of Alice (for $m= 1,\dots,M)$. Thus, the complex channel matrix between the IRS elements and all the sensors on Alice is given by $${\bf G}=[{\bf g}_{1},{\bf g}_{2},\dots,{\bf g}_{M}]\in \mathcal{C}^{K\times M}.$$ %with an additive white Gaussian noise (AGWN) matrix of $\boldsymbol{\Sigma} \in \mathcal{C}^{K\times M}.$ 
Invariably, considering the interaction between the $k$th IRS element and the $m$th sensor, we have that the incident signal in \eqref{rxu} can be expressed as 
\begin{multline}\label{rxk}
    g_{mk}[n]w_m[n]{\rm x}(t)[n]={\rm Re}\{w_m[n]\sqrt{c_{mk}[n]}s(t-\tau_{mk}[n])\\
    \times \exp{(j\omega_c(t-\tau_{mk}[n]))}\},
\end{multline}
where $\tau_{mk}[n]\coloneqq \frac{\|{\bf q}_{k}[n]-\Omega_m\|\hat{a}_{mk}}{c}$ defines the time delay between the $k$th IRS element and the $m$th antenna element and $g_{mk}$ presuming an exponential distribution with channel power gain, $c_{mk}[n]\coloneqq\frac{\rho_0\varsigma_{\rm G}}{\|{\bf q}_k[n]-\Omega_m\|^2}$ for $\rho_0$ and $\varsigma_{\rm G}$ representing the channel power gain at reference distance $d_0=1$m from the the centre of Alice and an exponentially distributed random variable with unit mean, respectively \cite{obinna_2020, uav_cooperative_jamming, uav_secured_com_JTTP}. Note that $\rho_0$ determines the quality of the channel which is then scaled randomly by $\varsigma_{\rm G}$ via the inverse distance square. Contrarily to \cite{Zarifi_2010}, we note that although the distance between the IRS elements and the distance between the sensor nodes are less than the distance between the UAV and Alice (that is $\|{\bf q}_k[n]-{\bf q}_{k\pm 1}[n]\|\ll\|{\bf q}_{\rm R}[n]-\Omega_{\rm A}\|$ and $\|\boldsymbol{\Omega}_m-\boldsymbol{\Omega}_{m\pm 1}\|\ll\|{\bf q}_{\rm R}[n]-\Omega_{\rm A}\|$), the independent sensor to IRS element variations cannot be ignored since the separation between sensors can be significant when deployed in multi-faceted environment. Nevertheless, we assume that the beam from the sensors is directed to the UAV carrying the IRS and not necessarily each IRS element, therefore the phase direction is towards the UAV. %This implies that the instantaneous location of the UAV within the $n$th sample is with reference to the center of Alice and not necessarily with respect to the individual sensors in Alice.
% the change in the channel power gain for each IRS element due to the $m$th sensor node in Alice would be negligible, hence, $c_{mk}[n]\approx c_{\rm AR}[n]\coloneqq\frac{\rho_0\varsigma}{\|{\bf q}_{\rm R}[n]-\Omega_{\rm A}\|^2}$. 
%Similarly, the time variation on each IRS element due to each sensor node would be seemingly equal such that $\tau_{mk}\approx \tau_{bk}[n]\coloneqq \frac{\|{\bf q}_{k}[n]-\Omega_{\rm A}\|\hat{a}_{bk}}{c}$. 
Equation \eqref{rxk} can then be modified for each $n\in \{1,\dots,N \}$ as 
\begin{multline}\label{rxk2}
    g_{mk}w_m{\rm x}(t)={\rm Re}\{w_m\sqrt{c_{m{\rm R}}}\exp{(-j\phi_{mk}\hat{a}_{m{\rm R}})}\\
    \times \delta\left(t-\frac{\phi_{mk}}{\omega_{\rm c}}\right)s(t)\exp{(j\omega_{\rm c}t)}\},
\end{multline}
where $\phi_{mk}\coloneqq \omega_{\rm c}\tau_{mk}$ characterizes the phase shift due to the $k$th IRS element location relative to the $m$th sensor. By implication of the distance between the IRS being far less than the distance between the UAV and Alice, the phase shift for each $n\in \{1,\dots,N\}$ can be approximated as 
$$\phi_{mk}\hat{a}_{m{\rm R}}\approx \underbrace{\frac{2\pi}{\lambda}\|{\bf q}_{\rm R}-\boldsymbol{\Omega}_m\|}_{\phi_{m{\rm R}}^a}+\underbrace{\frac{2\pi\hat{a}_{m{\rm R}}}{\lambda}\|{\bf q}_k-{\bf q}_{\rm R}\|\hat{a}_{Rk}}_{\phi_{mk}^b}.$$
It is imperative, then, that the phase shift is comprised of two distinct parts; based on i) the position of the UAV $(\phi_{m{\rm R}}^a)$ and ii) the $k$th IRS element response $(\phi_{mk}^b)$ to the signal from the $m$th sensor in Alice. In component form, the $k$th IRS element response phase can be written as \eqref{ph}
\begin{align}\label{ph}
    \phi_{mk}^b= [(k_{\rm x}-1)\bar{d}_{\rm x},(k_{\rm y}-1)\bar{d}_{\rm y},0]\cdot[\hat{a}_{m{\rm R}}^x,\hat{a}_{m{\rm R}}^y,\hat{a}_{m{\rm R}}^z]^T,
\end{align}
where $\bar{d}_{\rm x}=\frac{2\pi d_{\rm x}}{\lambda}$ and $\bar{d}_{\rm y}=\frac{2\pi d_{\rm y}}{\lambda}.$ It is easy to deduce that the IRS array response to all the sensor nodes can be represented with a $K\times M$ matrix, $\boldsymbol{\Phi}_{\rm G}$, whose elements are given in \eqref{ph}. The ${\rm rank}\{\boldsymbol{\Phi}_{\rm G}\}\geq 1$ depends on the value of $r$. $r>0$ increases the possibility of large distance between the $M$ sensors, thereby causing significant variations for the elements in $\boldsymbol{\Phi}_{\rm G}$. 

By extracting the complex channel coefficients from \eqref{rxk2} for each $n\in\{1, \dots, N\}$, the complex channel between the $k$th IRS element and the $m$th sensor node in Alice is thus presented as
\begin{align}\label{rxk3}
    g_{mk}=\sqrt{c_{m{\rm R}}}\exp{(-j\phi_{m{\rm R}}^a)}\exp{(-j \phi_{mk}^b)}\delta(t-\frac{\phi_{mk}}{\omega_{\rm c}}),
\end{align}
%where $\exp{(-j \phi_{mk}^b)}$ elucidates the $k$th IRS elements response to the incident signal from the $m$th sensor. %Generally, the combined IRS response to the incident signal from the $m$th sensor in Alice is ${\bf j}_{\rm B}\in \mathcal{C}^{K\times 1}$.
%\begin{multline}
%    {\bf j}_{\rm B}=[1,\dots,\exp{(-j(K_{\rm x}-1)\bar{d}_{\rm x}\hat{a}_{m{\rm R}}^x)}]^T\otimes\\
  %  [1,\dots,\exp{(-j(K_{\rm y}-1)\bar{d}_{\rm x}\hat{a}_{m{\rm R}}^y)}]^T,
%\end{multline}
%where $\bar{d}_{\rm x}=\frac{2\pi d_{\rm x}}{\lambda}$ and $\bar{d}_{\rm y}=\frac{2\pi d_{\rm y}}{\lambda}$. Therefore, the channel vector between all the IRS elements and the $m$th sensor can be presented as $${\bf g}_m=\sqrt{c_{m{\rm R}}}\exp{(-j\phi_{m{\rm R}}^a)}{\bf j}_{\rm B}\circ\delta(t-\frac{\phi_{m{\rm R}}^a+{\bf j}_{\rm B}}{\omega_{\rm c}}),$$ with ${\bf j}_{\rm B}$ uniquely showing the influence of the IRS formation on the incident signal.

Similarly, for each $n\in[1,\dots,N]$, the complex channel between the IRS and the ground nodes (that is ${\bf h}_{\rm i}=[h_{1i},\dots,h_{K{\rm i}}]^T\in \mathcal{C}^{K\times 1}, \forall{~{\rm i}\in\{{\rm B, E}\}}$), is obtained by updating the direction of the reflected signal to obtain 
\begin{align}\label{rxud}
    h_{k{\rm i}}=\sqrt{c_{\rm Ri}}\exp{(-j\phi_{\rm Ri}^a)}\exp{(-j \phi_{k{\rm i}}^b)}\delta(t-\frac{\phi_{k{\rm i}}}{\omega_{\rm c}}),
\end{align}
where $\phi_{k{\rm i}}^b= [(k_{\rm x}-1)\bar{d}_{\rm x},(k_{\rm y}-1)\bar{d}_{\rm y},0]\times[\hat{a}_{\rm Ri}^x,\hat{a}_{\rm Ri}^y,\hat{a}_{\rm Ri}^z]^T$. It is imperative that the generalized IRS array response to the ground node can be presented as a $K\times 1$ vector, ${\bf u}_{\rm i}$ with elements given as $\phi_{k{\rm i}}^b.$ Note that $c_{k{\rm i}}[n]\approx c_{\rm Ri}[n]\coloneqq\frac{\rho_0\varsigma_{\rm i}}{\|{\bf q}_{\rm R}[n]-\Omega_i\|^2},$ $\phi_{\rm Ri}^a=\frac{2\pi}{\lambda}\|\boldsymbol{\Omega}_{\rm i}-{\bf q}_{\rm R}\|$ and $\tau_{k{\rm i}}\coloneqq \frac{\|\boldsymbol{\Omega}_{\rm i}-{\bf q}_{\rm R}\|}{c}.$ 
%Given that the response from all the IRS elements to the reflected signal is
%\begin{multline}
 %   {\bf j}_{\rm i}=[1,\dots,\exp{(-j(K_{\rm x}-1)\bar{d}_{\rm x}\hat{a}_{\rm Ri}^x)}]^T\otimes\\
  %  [1,\dots,\exp{(-j(K_{\rm y}-1)\bar{d}_{\rm x}\hat{a}_{\rm Ri}^y)}]^T \sim \mathcal{C}^{K\times 1},
%\end{multline}
%the complex channel can be written as $${\bf h}_{\rm i}=\sqrt{c_{\rm Ri}}\exp{(-j\phi_{\rm Ri}^a)}{\bf j}_{\rm i}\circ\delta(t-\frac{\phi_{\rm Ri}^a+{\bf j}_{\rm i}}{\omega_{\rm c}}).$$ 
Therefore, having defined the channel parameters, the coherently received signal at the ground nodes (Bob and Eve)  during the $n$th sample is given by
\begin{align}\label{rxgnd}
    y_{\rm i}={\bf h}_{\rm i}^{\rm H}\boldsymbol{\Theta}{\bf G}{\bf w}{{\rm x}}(t)+\eta_{\rm i},
\end{align}
where ${\rm i}\in\{{\rm B, E}\}$, ${\bf w}=[w_1,\dots,w_M]^T$ is the beamforming weights of the $M$ sensors, $\boldsymbol{\Theta}={\rm diag}(\exp{(j\theta_1)},\exp{(j\theta_2)},\dots,\exp{(j\theta_K)})$ represents the vectorized reflection coefficients of the IRS elements and $\eta_{\rm i} \sim \mathcal{C}(0,\sigma_{\rm i}^2)~$%={\bf h}_{\rm i}^{\rm H}\boldsymbol{\Theta}\boldsymbol{\Sigma}{\bf 1}+\sum_{k=1}^Kn_{k{\rm i}}$ 
presents an independent and identically distributed (i.i.d.) additive white Gaussian noise (AWGN) at the corresponding receiver\footnote{By setting $\sigma_{\rm i}^2$ to high value, we ensure that the communication channel is noisy.}. % with zero mean and power of $\sigma_{\rm i}^2$.
Note that at the ground receiver nodes, (Bob and Eve), the reflected signals from all the IRS elements are superimposed coherently \cite{Cui_2019, Lu_2020}. It is intended that such coherent superimposition will maximize the received signal power at the Bob while limiting the signal power at Eve. %Therefore for any given ground receiver node, ${\rm i}\in\{{\rm B, E}\}$, we have:
%\begin{multline}
%    \chi_{mi}=\frac{\rho_0\varsigma\exp{(-j\frac{2\pi}{\lambda}(d_{\rm Ri}+d_{m{\rm R}}))}}{d_{\rm Ri}d_{m{\rm R}}}\\
%    \times\sum_{k_{\rm x}=1}^{K_{\rm x}}\sum_{k_{\rm y}=1}^{K_{\rm y}}\exp{(-j[(k_{\rm x}-1)\bar{d}_{\rm x}(\hat{a}_{\rm Ri}^x+\hat{a}_{m{\rm R}}^x)\\
%    +(k_{\rm y}-1)\bar{d}_{\rm y}(\hat{a}_{\rm Ri}^y+\hat{a}_{m{\rm R}}^y)])}\cdot \exp{(j\theta_{k_{\rm x},k_{\rm y}})},
%\end{multline}
%for ${\bf h}_{\rm i}^{\rm H}\boldsymbol{\Theta}{\bf G}=[\chi_{1i},\dots,\chi_{Mi}] \in \mathcal{C}^{1\times M}$, $d_{\rm Ri}=\|\boldsymbol{\Omega}_{\rm i}-{\bf q}_{\rm R}\|$ and $d_{m{\rm R}}=\|{\bf q}_{\rm R}-\boldsymbol{\Omega}_m\|.$

Our design objective is to maximize the secrecy rate of the communication by choosing the optimal beamforming weight, trajectory of the UAV and the reflection coefficients of the IRS. Accordingly, we formulate the following optimization problem:
\begin{subequations}\label{P1}
\begin{align}
   {\rm P1:}~~ \max_{{\bf w,Q},\boldsymbol{\Theta}}~& \frac{1}{N}\sum_{n=1}^N\left[\log_2\left(\frac{1+\gamma_{\rm B}[n]}{1+\gamma_{\rm E}[n]}\right)\right],\label{obj}\\
    {\rm s.t.}~& 0\leq {\theta}_{k}[n]\leq 2\pi,\label{cons_1}\\
    %{\rm s.t.}~& 0\leq\theta_{k_{\rm x},k_{\rm y}}\leq 2\pi\\
    ~&\|{\bf q}[n]-{\bf q}[n-1]\|^2\leq (Z\alpha)^2 \label{cons_2}\\
    ~&{\bf w}[n]^{\rm H}{\bf w}[n]\leq P \label{cons_3},
\end{align}
\end{subequations}
%%%%%%%%%%%%%%%%%%%%%%%%%%%%%%%%%%%%%%%%%%%%%%%%
where $\gamma_{\rm i}[n]=\frac{1}{\sigma_{\rm i}^2}|{\bf h}_{\rm i}^{\rm H}[n]\boldsymbol{\Theta}[n]{\bf G[n]w[n]}|^2,~ {{\rm i}\in\{{\rm B, E}\}}$ represents the signal to noise ratio at Bob and Eve. The constraint in \eqref{cons_1} ensures that the principal argument of the reflection coefficient from the $k$th IRS is maintained while \eqref{cons_2} limits the distance covered by the UAV between sampling points. Finally, \eqref{cons_3} constrains the power transmitted from the sensors.

\section{Proposed Solution} \label{sec_prop_sol}
Problem P1 is a non-convex multi-varriable optimization problem that is difficult to solve directly due to the inter-dependence of the varriables and the non-convexity of the objective function. Hence, we sub-divide the problem by creating distinct sub-optimal problems from P1 \cite{Cui_2019, Lu_2020, obinna_2020}. The idea is to solve the sub-optimal problems iteratively until a change in the objective value of problem P1 is insignificant. However, we see from the subsequent sections that by designing the reflection coefficient in terms of the main receiver, Bob, the sub-optimal problems arising from problem P1 can be solved using non-iterative means.

%%%%%%%%%%%%%%%%%%%%%%%%%%%%%%%%%%%%%%%%%%%%%%%%%%%%%
%{\bf Lemma 1:} 
%\[{\rm rank}\{\boldsymbol{\Phi}_{\rm G}\}  \left\{
%  \begin{array}{lr}
%    = 1 & r=0\\
%    > 1 & r> 0.
%  \end{array}
%\right.
%\]
%\textit{Proof:}......Not complete.....

%From Lemma 1, we note that for $r=0,$ implies that the sensors are located at a single point which cannot support collaborative beamforming. However, for $r>0$, the ${\rm rank}\{\boldsymbol{\Phi}_{\rm G}\}>1,$ the sensors are randomly distributed over a circular region. This implies that the IRS array response will differ significantly to the signal from each sensor node. This will invariably affect ${\bf G}$ and will act destructively at Eve.............
%%%%%%%%%%%%%%%%%%%%%%%%%%%%%%%%%%%%%%%%%%%%%%%%%%%%%%

\subsection{Solving for $\boldsymbol{\Theta}$}
Since the IRS elements can effectively reflect signals along the desired direction, we aim to design the optimal reflection coefficients such that the signals contribute to Bob's reception constructively. Since we assume that the channels from the IRS to Bob and Eve are independent of each other, it is sufficient to optimize the reflection coefficients based on Bob's channel only. Accordingly, we design the reflection coefficients $\boldsymbol{\Theta}$, for a given trajectory ${\bf Q}$, and beamforming vector ${\bf w}$, such that it is not influenced by Eve's channel condition. It is therefore, apparent that the optimal $\boldsymbol{\Theta}$ can be determined for maximizing Bob's SNR $\gamma_{\rm B}.$ This can be done by extracting the following sub-problem from the original problem P1: 
%${\bf h}_{\rm i}^{\rm H}\bar{\boldsymbol{\Theta}}{\bf G}=\boldsymbol{\theta}^T{\bf H}_{\rm i}{\bf G},~ \forall~i\in \{{\rm B,E}\},$ where ${\bf H}_{\rm i}$ is a diagonal matrix with element of ${\bf h}_{\rm i}.$ 
%\begin{subequations}\label{P21}
%\begin{align}
  % {\rm P2:}~~ \max_{\boldsymbol{\theta}}~& \left(\frac{1+{\boldsymbol{\theta}}^{\rm H}{\bf A}\boldsymbol{\theta}}{1+{\boldsymbol{\theta}}^{\rm H}{\bf B}\boldsymbol{\theta}}\right)\\
 %   {\rm s.t.}~& 0\leq \boldsynbol{\theta}_{k_{\rm x},k_{\rm y}} \leq 2\pi
%\end{align}
%\end{subequations}
\begin{subequations}\label{P2}
\begin{align}
   {\rm P2:}~~ \max_{\boldsymbol{\Theta}}~& \gamma_{\rm B}[n],\\
   {\rm s.t.}~& 0\leq {\theta}_{k}[n]\leq 2\pi.
    %{\rm s.t.}~& 0\leq\theta_{k_{\rm x},k_{\rm y}}[n]\leq 2\pi.%\\
    %&\forall{~n\in[1,\dots,N]}.\nonumber
\end{align}
\end{subequations}
The solution obtained from solving \eqref{P2} eventually maximizes the information rate received by Bob $(\log_2(1+\gamma_{\rm B}[n]))$. For each $n\in\{1,\dots,N\}$, the solution to problem P2 ensures the maximum objective value since the signals from the IRS elements are added constructively. This implies that in terms of the IRS elements, $\arg({\bf h}_{\rm B}^{\rm H}\boldsymbol{\Theta}{\bf G})=1.$ Following similar derivation as in \cite{Lu_2020, Sixian_2020}, it is easy to see that for each $n \in \{1,\dots,N\},$ the solution to problem P2 is given by
%\begin{multline}
 %   \theta_{k_{\rm x},k_{\rm y}}=\theta_{\rm com}-(k_{\rm x}-1)\bar{d}_{\rm x}(\hat{a}_{\rm RB}^{\rm x}+\hat{a}_{m{\rm R}}^x)\\
 %   -(k_{\rm y}-1)\bar{d}_{\rm y}(\hat{a}_{\rm RB}^{\rm y}+\hat{a}_{m{\rm R}}^y),
%\end{multline}
\begin{align}
    \boldsymbol{\theta}=\theta_{\rm com}+{\bf u}_{\rm B}+{\bf u}_{\rm G},\label{theta}
\end{align}
where $\boldsymbol{\theta}=[\theta_1,\dots,\theta_K]^T, ~{\bf u}_{\rm B}=[\phi_{1{\rm B}}^b,\dots,\phi_{K{\rm B}}^b]^T$ and ${\bf u}_{\rm G}$ is the maximum left singular vector corresponding to the rank-1 (low rank) approximation of $\boldsymbol{\Phi}_{\rm G}.$ $\theta_{\rm com}$ is an arbitrary phase common to all elements of the IRS. This phase allows for the cancellation of unscrupulous phase elements at the receiver arising from the direct link between the Alice and Bob \cite{Lu_2020, wu_2020}. However, since there is no direct path between the Alice and the ground receiver nodes in this work as described in Section \ref{sec2}, $\theta_{\rm com}$ can be set to zero without loss of generality. 

\subsection{Solving for ${\bf w}$} \label{sec_w}
We adopt two different design strategies for the beamforming weights ${\bf w},$ with known trajectory, ${\bf Q},$ and reflection coefficients, $\boldsymbol{\Theta}$. Scheme 1 does not consider the presence of the eavesdropper while Scheme 2 does.
\begin{enumerate}
    %\item If we consider only the first hop of transmission, it is known that the optimal ${\bf w}^*$ is an MRT towards the UAV given in \eqref{w} \cite{Zarifi_2010}.
    %\begin{align}\label{w}
    %    {\bf w}^*=\sqrt{\frac{P}{M}}{\bf a}_0,
    %\end{align}
    %where ${\bf a}_0=[\exp{(j\phi_{1R}^a)},\dots,\exp{(j\phi_{M{\rm R}}^a)}]$ represents the sensors response to transmitting to the UAV carrying the IRS. This solution ensures that the determination of the weights are independent  of the presence of Eve or the receiver (Bob).
    \item Since the IRS can direct signals to specific targets, we consider transmission to Bob only ignoring the presence of Eve. Then the problem P1 reduces to
     \begin{subequations}\label{P3a}
    \begin{align}
       {\rm P3a:}~~ \max_{\bf w}~& (1+{\bf w}^{\rm H}{\bf Aw}),\\
        {\rm s.t.}~& {\bf w}^{\rm H}{\bf w}\leq P,
    \end{align}
    \end{subequations}
    where ${\bf A}=\frac{1}{\sigma_{\rm B}^2}({\bf h}_{\rm B}^{\rm H}\boldsymbol{\Theta}{\bf G})^{\rm H}({\bf h}_{\rm B}^{\rm H}\boldsymbol{\Theta}{\bf G}).$ It is known that the optimal ${\bf w}^*=\sqrt{P}{\bf u}_{\rm max}$ is an MRT beamformer towards the UAV, where ${\bf u}_{\rm max}$ is the eigenvector corresponding to the maximum eigenvalue of the channel matrix $\frac{1}{\sigma_{\rm B}^2}({\bf h}_{\rm B}^{\rm H}\boldsymbol{\Theta}{\bf G})^{\rm H}({\bf h}_{\rm B}^{\rm H}\boldsymbol{\Theta}{\bf G})$ \cite{Lu_2020, Zarifi_2010}. This solution ensures that the determination of the weights are independent  of the presence of Eve. \label{scheme_1_sec}
    
    \item Taking the eavesdropper's information into consideration, we reformulate problem P1 as
    \begin{subequations}\label{P3b}
    \begin{align}
       {\rm P3b:}~~ \max_{\bf w}~& \bigg(\frac{1+{\bf w}^{\rm H}{\bf Aw}}{1+{\bf w}^{\rm H}{\bf Bw}}\bigg),\\
        {\rm s.t.}~& {\bf w}^{\rm H}{\bf w}\leq P,
    \end{align}
    \end{subequations}
    where ${\bf A}=\frac{1}{\sigma_{\rm B}^2}({\bf h}_{\rm B}^{\rm H}\boldsymbol{\Theta}{\bf G})^{\rm H}({\bf h}_{\rm B}^{\rm H}\boldsymbol{\Theta}{\bf G}),~ {\bf B}=\frac{1}{\sigma_{\rm E}^2}({\bf h}_{\rm E}^{\rm H}\boldsymbol{\Theta}{\bf G})^{\rm H}({\bf h}_{\rm E}^{\rm H}\boldsymbol{\Theta}{\bf G}).$
    This implies that by considering the presence of both Bob and Eve, the optimal ${\bf w}^*=\sqrt{P}{\bf u}_{\rm max}$, where ${\bf u}_{\rm max}$ is the eigenvector corresponding to the maximum eigenvalue of the matrix ${({\bf B}+\frac{1}{P}{\bf I}_M)}^{-1}({\bf A}+\frac{1}{P}{\bf I}_M)$ \cite{Cui_2019}. \label{scheme_2_sec}
\end{enumerate}
In this work, we examine the performance characteristics of both schemes in order to determine their impact on the average secrecy rate.

\subsection{Solving for {\bf Q}}
Now, following the sub-optimal solutions for ${\bf w},~\boldsymbol{\Theta},$ obtained in the preceding subsections,  we define Proposition 1 at Bob and Eve for each $n\in\{1,\dots,N\}.$

\textbf{Proposition 1:} $|{\bf h}_{\rm i}^{\rm H}\boldsymbol{\Theta}{\bf G}|\leq |\boldsymbol{\chi}_{{\rm i}}|$
%${\bf h}_{\rm B}\boldsymbol{\Theta}{\bf G}$ and ${\bf h}_{\rm E}\boldsymbol{\Theta}{\bf G}$ 
where the elements of $\boldsymbol{\chi}_{{\rm i}}$ are presented in \eqref{htG_bs} and \eqref{htG_e1} for ${{\rm i}\in\{{\rm B, E}\}},$ respectively: 
\begin{subequations}\label{kappa}
\begin{align}
    \chi_{m{\rm B}}=K\sqrt{c_{{\rm RB}}c_{m{\rm R}}}\exp{(-j(\phi_{m{\rm R}}^a+\phi_{{\rm RB}}^a-\theta_{{\rm com}}))},\label{htG_bs}
\end{align}
\begin{multline} 
    \chi_{m{\rm E}}=\sqrt{c_{{\rm RE}}c_{m{\rm R}}}\exp\left(-j\left(\phi_{{\rm RE}}^a+\phi_{m{\rm R}}^a-\theta_{{\rm com}}\right)\right)\\
    \times \left(\sum_{k_{{\rm x}}=1}^{K_{{\rm x}}}\sum_{k_{{\rm y}}=1}^{K_{{\rm y}}}\exp\bigg(j\left[(k_{{\rm x}}-1)\bar{d}_{{\rm x}}(\hat{a}_{{\rm RE}}^{{\rm x}}+\hat{a}_{{\rm RB}}^{{\rm x}}-\hat{a}_{m{\rm R}}^{{\rm x}})\right.\right.\\
    \left.\left. +(k_{{\rm y}}-1)\bar{d}_{{\rm y}}(\hat{a}_{{\rm RE}}^{{\rm y}}+\hat{a}_{{\rm RB}}^{{\rm y}}-\hat{a}_{m{\rm R}}^{{\rm y}})+{u}_{\rm G}^{k_{\rm x},k_{\rm y}}\right]\bigg)\right), \label{htG_e1}
\end{multline}
\end{subequations}
where ${u}_{\rm G}^{k_{\rm x},k_{\rm y}}$ is the $k$th element of ${u}_{\rm G}$ and $\boldsymbol{\chi}_{{\rm i}}=[\chi_{1{\rm i}},\dots,\chi_{M{\rm i}}].$ 

\textit{Proof:} By substituting and simplifying the expressions for ${\bf h}_{\rm i}, ~\boldsymbol{\Theta}, {\rm ~and,~} {\bf G}$ given in equations \eqref{rxud}, \eqref{theta} and \eqref{rxk3} respectively, it is easy to see that ${\bf u}_{\rm G}$ in \eqref{theta} is designed as a rank-1 approximation to cancel out the variations of the columns of matrix $\boldsymbol{\Phi}_{\rm G}.$ However, we know that the rank-1 approximation error of ${\bf u}_{\rm G}$ increases as $r>0$ for the $m$th sensor, hence $|{\bf u}_{\rm G}-\boldsymbol{\Phi}_{\rm G}(:,m)|\geq 0$. The upper bound in Proposition 1 is thus determined considering that $|{\bf u}_{\rm G}-\boldsymbol{\Phi}_{\rm G}(:,m)|= 0$. Therefore it is apparent to state that $|\boldsymbol{\chi}_{{\rm B}}|$ and $|\boldsymbol{\chi}_{{\rm E}}|$ are the upper bounds of $|{\bf h}_{\rm B}\boldsymbol{\Theta}{\bf G}|$ and $|{\bf h}_{\rm E}\boldsymbol{\Theta}{\bf G}|$, respectively, thereby completing the proof of Proposition 1. \hfill $\blacksquare$

Following Proposition 1, we can now determine the maximum SNR values at Bob and Eve as
\begin{subequations}\label{SNR}
\begin{align}
    \gamma_{\rm B}\leq \sum_{m=1}^M\frac{\bar{P}(\rho_0\varsigma_{\rm B} K)^2}{d_{\rm RB}^2d_{\rm mR}^2},\\
    \gamma_{\rm E}\leq \sum_{m=1}^M\frac{\bar{P}(\rho_0\varsigma_{\rm E} \zeta)^2}{d_{\rm RE}^2d_{\rm mR}^2},
\end{align}
\end{subequations}
respectively, where $\bar{P}=\frac{P}{\sigma_{i}^2}$ and 
\begin{multline*}
    \zeta=\sum_{k_{{\rm x}}=1}^{K_{{\rm x}}}\sum_{k_{{\rm y}}=1}^{K_{{\rm y}}}\exp\bigg(j\bigg[(k_{{\rm x}}-1)\bar{d}_{{\rm x}}(\hat{a}_{{\rm RE}}^{{\rm x}}+\hat{a}_{{\rm RB}}^{{\rm x}}-\hat{a}_{m{\rm R}}^{{\rm x}})\\
    +(k_{{\rm y}}-1)\bar{d}_{{\rm y}}(\hat{a}_{{\rm RE}}^{{\rm y}}+\hat{a}_{{\rm RB}}^{{\rm y}}-\hat{a}_{m{\rm R}}^{{\rm y}})+{u}_{\rm G}^{k_{\rm x},k_{\rm y}}\bigg]\bigg)\overset{(a)}{\leq} K.
\end{multline*}
Note that the bounds in \eqref{SNR} invariably define the bounds of the average secrecy rate. 
%maximum average secrecy rate attainable given that $r>0.$ During the design of IoT systems, this relation is necessary as a guarantee to secrecy. 
The equality in $(a)$ represents the worst-case scenario and arises when the channel of Bob and Eve are highly correlated. This may occur in the unlucky event when Eve is located at the exact position of Bob (e.g. an application in the device of Bob becoming the potential Eve).

Now, to obtain the trajectory of the UAV for known IRS reflection coefficients, $(\boldsymbol{\Theta})$ and beamforming vectors, ${\bf w}$, problem P1 is reformulated as problem P4: 

\begin{subequations}\label{P4}
\begin{align}
   {\rm P4:}~~~ \max_{{\bf Q}}~& \sum_{n=1}^N\left[\log_2\left(\frac{1+\gamma_{\rm B}[n]}{1+\gamma_{\rm E}[n]}\right)\right]\\
    {\rm s.t.}~&\|{\bf q}[n]-{\bf q}[n-1]\|^2\leq (Z\alpha)^2.
\end{align}
\end{subequations}
\comment{
\begin{subequations}\label{P5}
\begin{align}
   {\rm P5:}~~~ \max_{{\bf Q}}~& \sum_{n=1}^N\left[\log_2\left(\frac{1+\gamma_{\rm B}[n]}{\beta [n]}\right)\right],\\
    {\rm s.t.}~& 1+\gamma_{\rm E}[n] \leq \beta [n]\\
    &\|{\bf q}[n]-{\bf q}[n-1]\|^2\leq (Z\alpha)^2. \label{tracons}
\end{align}
\end{subequations}
}
Note that the problem P4 is non-convex due to the fractional objective. The problem (P4) can be solved by introducing a slack variable $\beta$ limiting the maximum achievable rate by Eve as shown in \eqref{P5}. Furthermore, considering that the distance between the sensors in Alice is very small compared to the distance, between Alice and the UAV-carried IRS, we can assume for simplification that the UAV trajectory is determined in respect to Alice rather than the individual sensors, $(d_{\rm mR}\approx d_{\rm AR})$. Therefore, we reformulate the trajectory problem to Problem P5.
\begin{subequations}\label{P5}
\begin{align}
   {\rm P5:}~~~ \max_{{\bf Q}, \beta}~& \sum_{n=1}^N\left[\log_2\left(1+\frac{\bigg(\frac{\bar{P}M(\rho_0\varsigma_{\rm B} K)^2}{d_{\rm RB}^2d_{\rm AR}^2}\bigg)[n]}{\beta [n]}\right)\right],\\
    {\rm s.t.}~& 1+\bigg(\frac{\bar{P}M(\rho_0\varsigma_{\rm E} K)^2}{d_{\rm RB}^2d_{\rm AR}^2}\bigg)[n] \leq \beta [n]\\
    &\|{\bf q}[n]-{\bf q}[n-1]\|^2\leq (Z\alpha)^2. \label{tracons6}
\end{align}
\end{subequations}
Problem P5 can be solved using Karush-Kuhn-Tucker (KKT) conditions to obtain the optimal trajectory of the UAV as defined in Proposition 2. A detailed proof is relegated to Appendix~\ref{Appendix_A}.

{\bf Proposition 2:} Given the maximum achievable rate at Eve is $\beta$, the optimal location of the UAV during the $n$th sample $(n\in[1,\dots,N])$ can be obtained by solving
\begin{align} \label{qsol}
    q_{\rm x}^2[n]+q_{\rm y}^2[n]= (\varepsilon[n]\|\Omega_{\rm A}\|)^2-H^2.
\end{align}
\textit{Proof:} See Appendix \ref{Appendix_A}. \hfill $\blacksquare$

\begin{figure*}[ht!]
\begin{align} \label{va}
    \varepsilon[n]=\frac{\frac{1}{3\sqrt[3]{2}}\sqrt[3]{2b^3+3\sqrt{3}\sqrt{-4b^3d-b^2c^2+18bcd+4c^3+27d^2}-9bc-27d}-\sqrt[3]{2}(3c-b^2)}{3\sqrt[3]{2b^3+3\sqrt{3}\sqrt{-4b^3d-b^2c^2+18bcd+4c^3+27d^2}-9bc-27d}+\frac{b}{3}}, 
\end{align}
where $b=\frac{\|\boldsymbol{\Omega}_{\rm B}\|}{2\|\boldsymbol{\Omega}_{\rm A}\|}+2\frac{\|{\bf q}[n-1]\|}{\|\boldsymbol{\Omega}_{\rm A}\|}+\frac{1}{2}$, $c=\frac{\|\boldsymbol{\Omega}_{\rm B}\|\|{\bf q}[n-1]\|+\|{\bf q}[n-1]\|^2}{\|\boldsymbol{\Omega}_{\rm A}\|^2}-\frac{\|{\bf q}[n-1]\|}{\|\boldsymbol{\Omega}_{\rm A}\|}$, $d=\frac{\|\boldsymbol{\Omega}_{\rm B}\|}{2\|\boldsymbol{\Omega}_{\rm A}\|}\frac{\|{\bf q}[n-1]\|^2}{\|\boldsymbol{\Omega}_{\rm A}\|^2}+\frac{\|{\bf q}[n-1]\|^2}{2\|\boldsymbol{\Omega}_{\rm A}\|^2}-\frac{(Z\alpha)^2}{2\|\boldsymbol{\Omega}_{\rm A}\|^2}$ \\
\_\_\_\_\_\_\_\_\_\_\_\_\_\_\_\_\_\_\_\_\_\_\_\_\_\_\_\_\_\_\_\_\_\_\_\_\_\_\_\_\_\_\_\_\_\_\_\_\_\_\_\_\_\_\_\_\_\_\_\_\_\_\_\_\_\_\_\_\_\_\_\_\_\_\_\_\_\_\_\_\_\_\_\_\_\_\_\_\_\_\_\_\_\_\_\_\_\_\_\_\_\_
\end{figure*}

The closed-form expression for $\varepsilon$ is given in \eqref{va} (at the top of the next page). The solution to \eqref{qsol} can easily be obtained by a linear search algorithm that seeks for pairs of points that satisfy the trajectory constraint in \eqref{tracons6}. We note that the trajectory is related to the solution of \eqref{qsol} by ${\bf Q}=\{{\bf q}[n]=[q_{\rm x}[n],q_{\rm y}[n], H]^T, n\in \{1,\dots,N\}\}.$ It can be deduced from Proposition 2 that the trajectory of the UAV is not dependent on the knowledge of the rate received at Eve or Bob but on the exact location of Bob, Alice and the distance covered by the UAV during the $n$th sample. This ensures that the rate regulation (varying $\beta$) for Eve and the knowledge of the location of Eve are insignificant in determining all the possible locations of the UAV for $n\in\{1,\dots,N\}.$ However, while conducting the linear search to obtain the optimal location among the possible locations, the knowledge of the location of Eve influences the choice leading to the trajectory of the UAV as obtained in Fig.~\ref{tra}. The overall procedure is summarised in Algorithm~\ref{algo1}.
%%%%%%%%%%%%%%%%%%%%%%%%%%%%%%%%%%%%%%%%%%%%%%%%%%%%
%%%%%%%%%%%%%%%%%%%%%%%%%%%%%%%%%%%%%%%%%%%%
\begin{algorithm} [!ht]
    \caption{Algorithm for solving $\boldsymbol{\Theta},~{\bf w},~\textrm{and}~{\bf Q}$}
  \begin{algorithmic}[1]\label{algo1}
%    \STATE Initialize ${\bf w}^0~\textrm{and}~ {\bf q}^0$ such that the constraints in \eqref{cons_3} and \eqref{cons_2} are respectively satisfied. Then set the objective value defined in \eqref{obj} as $R_s^0=0$
    %\STATE $m \leftarrow 1.$
    %\STATE \textbf{repeat}
    \STATE Solve \eqref{qsol} and update ${\bf q}$.
    \STATE Using the grid cell, ${\bf q},$ $d_{\rm x},~{\rm and}~ d_{\rm y},$ compute the locations of the IRS elements, ${\bf Q}.$
    \STATE Determine the channel impulse responses using the definitions in \eqref{rxk3} and \eqref{rxud}.
    \STATE For each $n \in \{1,\dots,N\},$ solve \eqref{theta} to obtain $\boldsymbol{\Theta}.$
    \STATE Compute and update ${\bf w}$ with solutions described in \ref{sec_w}. % computed $\boldsymbol{\Theta}$ and channel impulse responses with .
    
    \STATE Compute $R_s$ as defined in \eqref{obj}.
    %\STATE Compute $e=\bigg|\frac{R_s^{new}-R_s^{old}}{R_s^{new}}\bigg |$.
    %\STATE $m \leftarrow m + 1.$
    
    %\STATE \textbf{until} {$e< \theta$ OR $m\geq m_{max}$.}
    \STATE \textbf{Output:} $\boldsymbol {\Theta},~{\bf w}~\textrm{and}~\boldsymbol {\rm Q}$.
  \end{algorithmic}
\end{algorithm}
%%%%%%%%%%%%%%%%%%%%%%%%%%%%%%%%%%%%%%%%%%%

\section{Results and Discussions} \label{sec3}

In this section, we evaluate the performance of the proposed algorithm via numerical simulations and compare with baseline schemes. 
The parametric settings of the simulation environment are given in Table \ref{table_par} except explicitly stated.
%: $M=4,~\boldsymbol{\Omega}_{\rm A}=[0,-100,0]^T, ~\boldsymbol{\Omega}_{\rm E}=[-100,50,0]^T~{\rm (Uncorrelated)}, ~\boldsymbol{\Omega}_{\rm E}=[75,100,0]^T~{\rm (Correlated)}, ~\boldsymbol{\Omega}_{\rm B}=[80,100,0]^T, ~\boldsymbol{\Omega}_{\rm fixIRS}=[100,80,0]^T, ~H=100{\rm m}, ~Z=3{\rm m/s}, ~\rho_0=120{\rm dBm}, ~\sigma_{\rm B}^2=\sigma_{\rm E}^2=30{\rm dBm}\footnote{By setting high noise power, we ensure that the communication channel is noisy.}, ~q_0=[-100,100,H]^T, ~f_{\rm c}=900{\rm MHz}, ~z_{\rm x}=z_{\rm y}=4.$ 
\begin{table}[!ht]
\begin{center}
\caption{Simulation Parameters}\label{table_par}
    \begin{tabular}{ | p{2.7cm} | l | l |}
   
    \hline
    \textbf{Simulation parameter} & \textbf{Symbol} & \textbf{Value} \\ \hline
    Number of sensors & $M$ & $4$ \\ \hline
    Center of Alice & $\boldsymbol{\Omega}_{\rm A}$ & $[0,-100,0]^T$ \\ \hline
    Bob location & $\boldsymbol{\Omega}_{\rm B}$ & $[80,100,0]^T$ \\ \hline
    Eve location & $\boldsymbol{\Omega}_{\rm E}$ & $[-100,50,0]^T~({\rm Uncorrelated})$\\
    & & $[75,100,0]^T~({\rm Correlated})$\\ \hline
    %Eve location (Correlated)& $\boldsymbol{\Omega}_{\rm E}$ & $[75,100,0]^T~({\rm Correlated})$ \\ \hline
    Fixed IRS location & $\boldsymbol{\Omega}_{\rm fixIRS}$ & $[80,100,0]^T$ \\ \hline
    Initial UAV location & $\boldsymbol{q}_o$ & $[-100,100,H]^T$ \\ \hline
    UAV height & $H$ & $100$m \\ \hline
    UAV time of flight & $T$ & $300$s \\ \hline
    Velocity per sample & $Z$ & $3$m/s \\ \hline
    Duration per sample & $\alpha$ & $0.5$s \\ \hline
    Transmission frequency & $f$ & $900$ MHz\\ \hline
    Number of IRS elements & $K$ & $16$\\ \hline
    IRS separation & $d_{\rm x}=d_{\rm y}$ & $\frac{\lambda}{4}$ \\ \hline
    Noise power & $\sigma_{\rm B}^2=\sigma_{\rm E}^2$ & $30$dBm \\ \hline
    Signal-to-noise ratio & $\rho_o$ & $120$dBm (strong),\\ 
    & & $60$dBm (weak) \\ 
   \hline  
    \end{tabular}
\end{center}
\end{table}
We use the process in Algorithm~\ref{algo1} to optimize the parameters with initial values satisfying respective constraints. 
%This process continues until the error ($e$) between steps is less than $\theta$ (where $\theta=10^{-5}$) or the maximum number of iterations is reached (where $m_{max}=200$). 
The legend of the figures describe the various scenarios implemented as:
\begin{enumerate}
    \item Scheme 1: Refers to the UAV-carried IRS scenario where the beamforming weights are obtained with the expression in Section \ref{sec_w}.\ref{scheme_1_sec}.
    \item Scheme 2: Refers to the UAV-carried IRS scenario where the beamforming weights are obtained with the expression in Section \ref{sec_w}.\ref{scheme_2_sec}.
    \item Fixed: Refers to the Algorithm 1 given in \cite{Cui_2019}. To adapt the algorithm to the scenario described herein, we replaced the structured transmit antenna at the AP with a sensor network and set the direct link between Alice and Bob/Eve to $0$.
\end{enumerate}

%%%%%%%%%%%%%%%%%FIGURES%%%%%%%%%%%%%%%%%%%%%%%%%
%%%%%%%%%%%%%%%%%%%%%%%%%%%%%%%%%%%%%%%%%%%%%%%%%%
%Trajectory
%\renewcommand\thesubfigure{\roman{subfigure}}
\begin{figure*}[tb]
\centering
\begin{subfigure}{.48\textwidth}
  \centering
  \includegraphics[width=\linewidth]{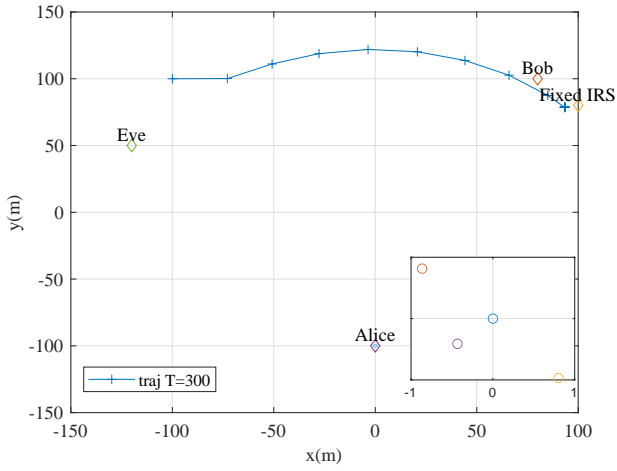}%[width=.4\linewidth]{image1}
  \caption{Eve's location is far away from Bob}
  \label{tra1}
\end{subfigure}
\begin{subfigure}{.48\textwidth}
  \centering
  \includegraphics[width=\linewidth]{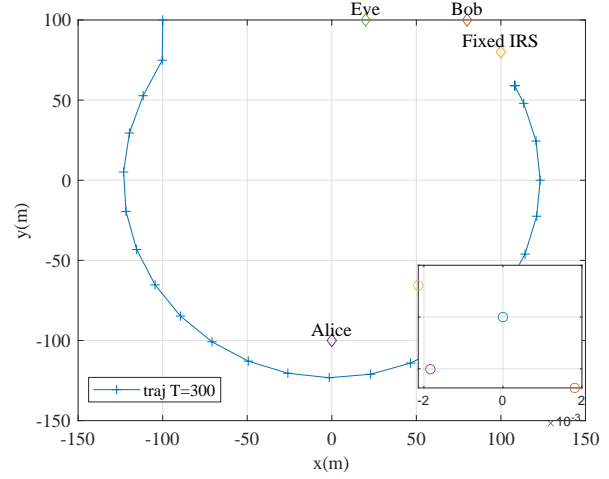}%[width=.4\linewidth]{image1}
  \caption{Eve's location is close to Bob}
  \label{tra2}
\end{subfigure}
\caption{UAV trajectory for different locations of Eve.}
\label{tra}
\end{figure*}
%%%%%%%%%%%%%%%%%%%%%%%%%%%%%%%%%%%%%%%%%%%%%%%%%%
%Time of flight
\begin{figure}[tb]
\centering
\includegraphics[width=1.0\linewidth]{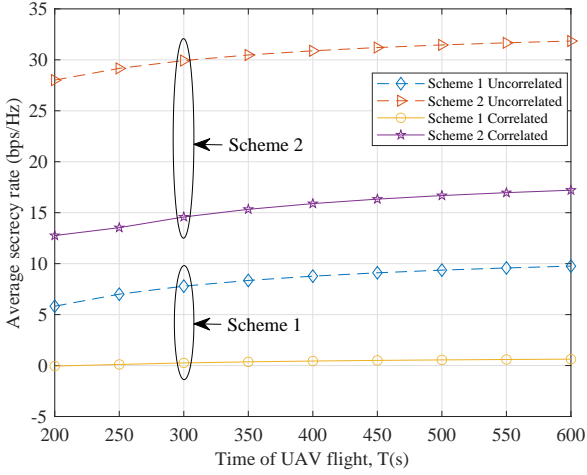}
\caption{Average secrecy versus time of flight (T) for different transmit power (dBm) for $K=16,~r=1{\rm m} {\rm ~and~} P=1{\rm dBm}$}
\label{tof}
\end{figure}
%%%%%%%%%%%%%%%%%%%%%%%%%%%%%%%%%%%%%%%%
%Transmit power
\begin{figure*}[tb]
\centering
\begin{subfigure}{.48\textwidth}
  \centering
  \includegraphics[width=\linewidth]{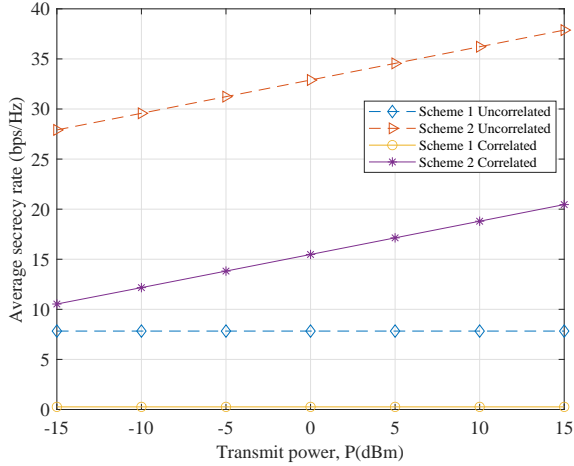}%[width=.4\linewidth]{image1}
  \caption{Strong channel quality for both Bob and Eve $\rho_0=120$dBm}
  \label{P_dBa}
\end{subfigure}
\begin{subfigure}{.48\textwidth}
  \centering
  \includegraphics[width=\linewidth]{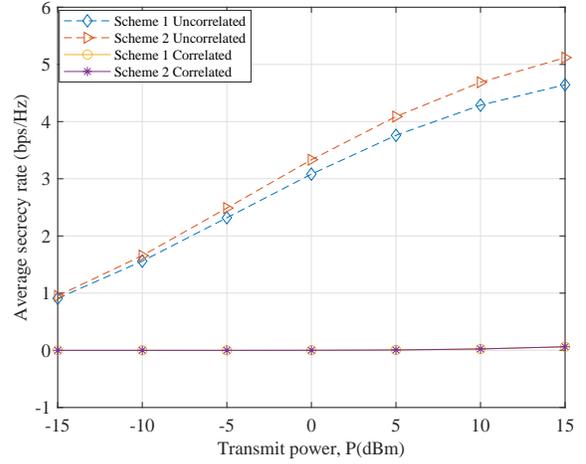}%[width=.4\linewidth]{image1}
  \caption{Weak channel quality for both Bob and Eve $\rho_0=60$dBm}
  \label{P_dBb}
\end{subfigure}
\caption{Average secrecy rate versus transmit power for $K=16,~r=1{\rm m} {\rm ~and~} T=300{\rm s}$.}
\label{P_dB}
\end{figure*}
%%%%%%%%%%%%%%%%%%%%%%%%%%%%%%%%%%%%%%%%%
%distance between Bob and Eve
\begin{figure*}[tb]
\centering
\begin{subfigure}{.48\textwidth}
  \centering
  \includegraphics[width=\linewidth]{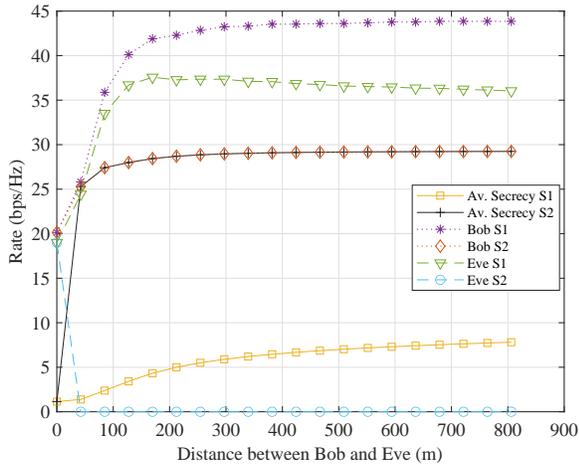}%[width=.4\linewidth]{image1}
  \caption{Strong channel quality for both Bob and Eve $\rho_0=120$dBm}
  \label{corr1}
\end{subfigure}
\begin{subfigure}{.48\textwidth}
  \centering
  \includegraphics[width=\linewidth]{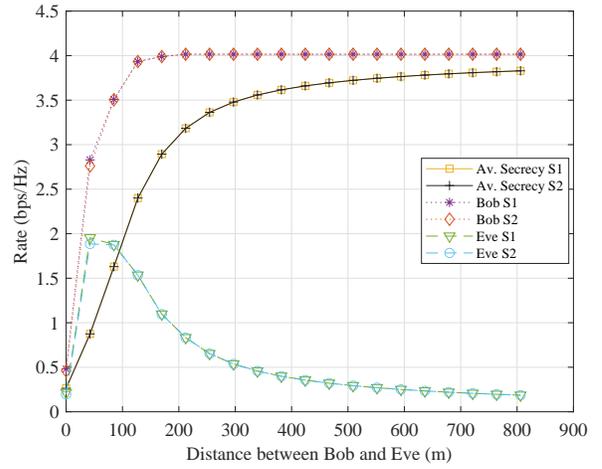}%[width=.4\linewidth]{image1}
  \caption{Weak channel quality for both Bob and Eve $\rho_0=60$dBm}
  \label{corr2}
\end{subfigure}
\caption{Average secrecy rate versus distance between Bob and Eve for $K=16,~r=1{\rm m},~P=1{\rm dBm} {\rm ~and~} T=300{\rm s}$.}
\label{corr}
\end{figure*}
%%%%%%%%%%%%%%%%%%%%%%%%%%%%%%%%%%%%%%%%%%
%Number of IRS
\begin{figure*}[tb]
\centering
\begin{subfigure}{.48\textwidth}
  \centering
  \includegraphics[width=\linewidth]{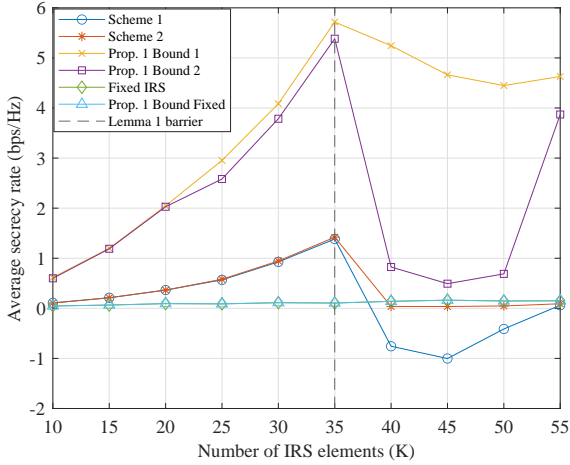}%[width=.4\linewidth]{image1}
  \caption{Weak channel quality for both Bob and Eve $\rho_0=60$dBm}
  \label{K_on_Eve1}
\end{subfigure}
\begin{subfigure}{.48\textwidth}
  \centering
  \includegraphics[width=\linewidth]{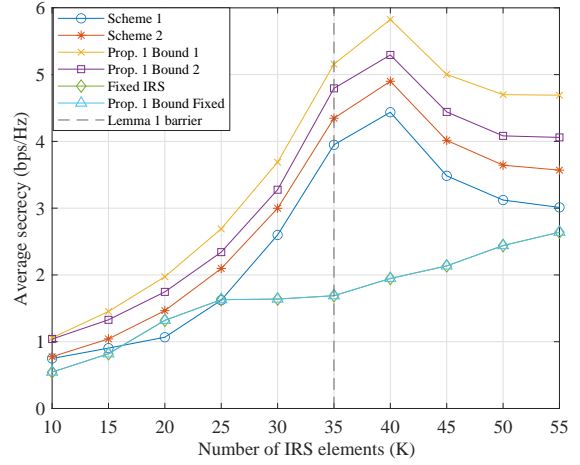}%[width=.4\linewidth]{image1}
  \caption{Strong channel quality for both Bob and Eve $\rho_0=120$dBm}
  \label{K_on_Eve2}
\end{subfigure}
\caption{Influence of the number of IRS on Average secrecy rate at possible formation of Eve and Bob channel $(r=1{\rm m} ~{\rm and}~ P=1{\rm dBm}, ~T=300s$).}
\label{K_on_Eve}
\end{figure*}
%%%%%%%%%%%%%%%%%%%%%%%%%%%%%%%%%%%%%%%%%%%%
%Radius
\begin{figure}[tb]
\centering
\includegraphics[width=1.0\linewidth]{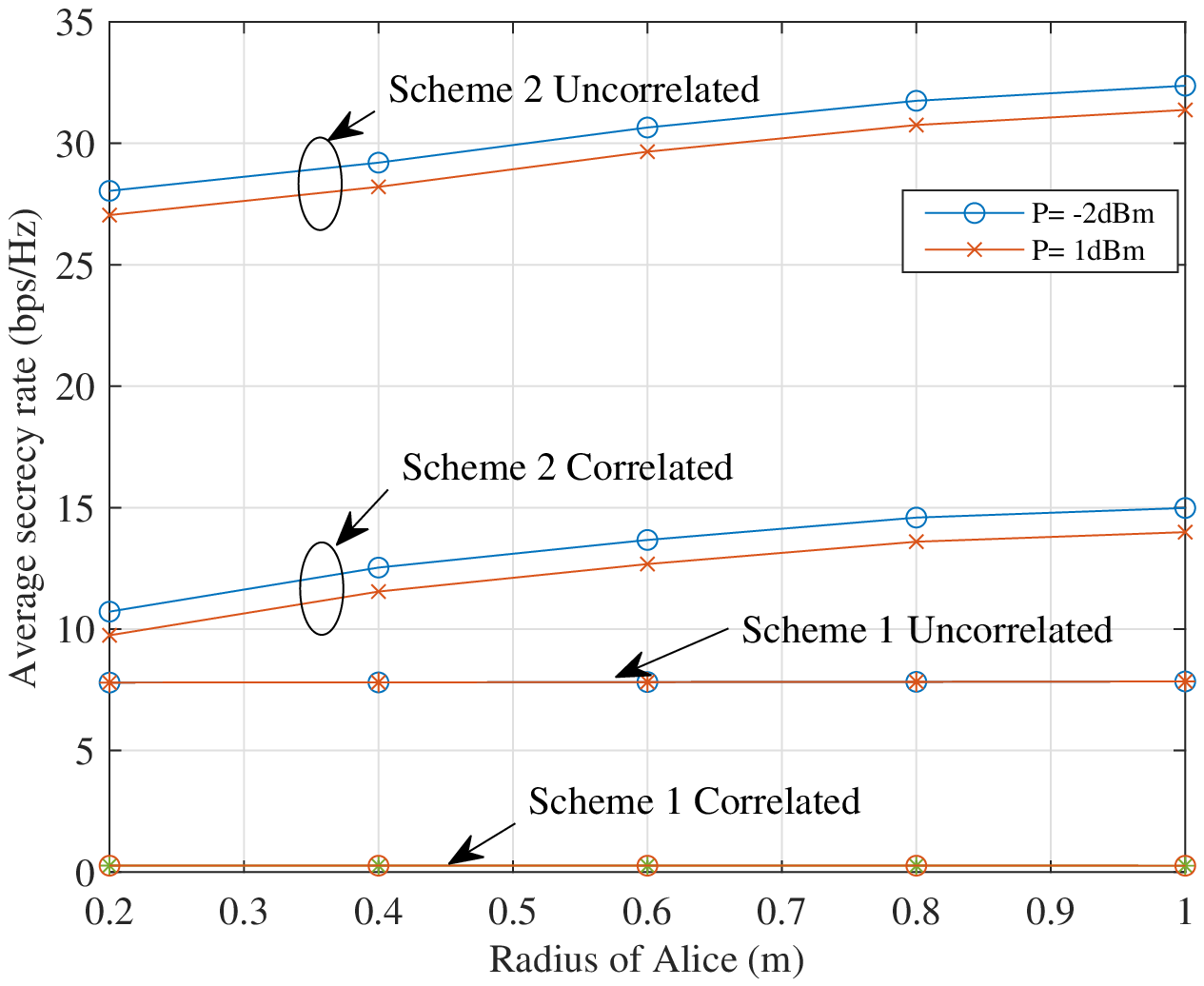}
\caption{Average secrecy rate versus Radius of sensor location for $K=16~{\rm and}~T=300s$}
\label{radius}
\end{figure}
%%%%%%%%%%%%%%%%%%%%%%%%%%%%%%%%%%%%%%%%%%%

Figure \ref{tra} presents the trajectory of the UAV as we change the location of Eve. We consider when Eve location is far away from Bob and when it is close to Bob in the sub-figures. The UAV attempts to find paths that are as far from Eve as possible while maintaining reasonable distance between Alice and the Bob to ensure the transmitted signals are received and reflected. When best safe distance is obtained, the UAV hovers around that location until the end of the simulation. This behaviour of the UAV is similar for different scheduled flight times. Intuitively, since the IRS is passive, for optimal performance, the distance traveled by the reflected signal is required to be small while maintaining LoS with Alice. The active beamforming at the Alice ensures that the transmitted signal are directed to the IRS on the UAV and can adjust the transmitted power where necessary. This intuitive trajectory of the UAV collaborates the conclusion in \cite{Emil_2020} having shown that the received signal power at Bob is proportional to the square of the IRS area and inverse square propagation distance, $\frac{1}{(d_{\rm AR}d_{\rm RB})^2}.$ Therefore, the optimal IRS placement should aim to minimize $d_{\rm AR}d_{\rm RB}$ as obtained via the UAV. Furthermore, the position of Eve determines the trajectory while the position of the main receiver (Bob) determines the the optimal location for the UAV-carried IRS system. Following the trajectory presented in Fig. \ref{tra}, it is observed in Fig. \ref{tof} that the longer the UAV flies with the IRS for a given communication, the better the average secrecy rate for both beamforming weight schemes under consideration. It has been established in \cite{Lu_2020} that for aerial IRS, the SNR increases with higher transmit power. However, due to the IRS, we showed that the SNR for Eve declines leading to an increase in the average secrecy rate as observed in the rate of the Eve in Fig. \ref{corr}. Similar performance was observed in the fixed IRS scenario as reported in \cite{Cui_2019}. Figure \ref{tof} also provides an insight that scheme 2 performs better than scheme 1 when the channels of Bob and Eve are correlated and uncorrelated. %However, this positive performance in terms of average secrecy is insignificant after 450 seconds giving the time the UAV stabilizes at the optimal location.

In Fig. \ref{P_dB}, the impact of transmit power on the average secrecy rate of the system was presented. By comparing the sub-figures, scheme 2 out-performs scheme 1 when the channel quality of Bob and Eve are strong (represented by different values of $\rho_0$). Similar assertion is presented in Fig. \ref{corr} while examining the influence of distance between Bob and Eve representing correlation are studied. %It is clear from the rate presentation in Fig. \ref{corr} that since 
In Fig. \ref{corr1} and Fig. \ref{P_dBa}, we note that scheme 2 was designed with the perfect knowledge of Eve, therefore, the information rate received by Eve is only maximum when the correlation\footnote{Since the channels in this paper were modeled as a function of distance, the distance between Bob and Eve $={\|\boldsymbol{\Omega}_{\rm B}-\boldsymbol{\Omega}_{\rm E}\|}$ examines the similarity between Bob and Eve in terms of their proximity. Lower values indicate highly correlation while higher values indicate highly uncorrelated \cite{Si_2019, sarka_2012}.} between the channels of Bob and Eve is highest and declines rapidly as the distance between Bob and Eve increases. A combination of the benefits of the beamforming weights in scheme 2 and the optimized reflection coefficients easily cause its average secrecy rate to the upper bound, which is the rate of Bob. Nevertheless, since the beamforming weights for scheme 1 was designed ignorant of Eve, the information rate received at Eve is compromised only by the influence of the reflection coefficients, hence its average secrecy rate is always below its upper bound. In contrast to these observations, Fig. \ref{corr2} and Fig. \ref{P_dBb}, elucidates that the performance in terms of both scheme 1 and 2 are similar when the channel quality of Bob and Eve are poor. This is an interesting result as it provides reasonable justification deploying scheme 1 especially when the exact location of Eve is unknown under noisy channel conditions. 

Furthermore, with the 2D separation between the elements of the IRS system in horizontal and vertical directions as $d_{\rm x} {~{\rm and}~} d_{\rm y}$ respectively, we define Lemma 1 to determine the maximum number of IRS elements $(K)$ to guarantee the average secrecy rate defined in the objective of problem P1. Lemma 1 is a consequence of \cite{Emil_2020} on noisy channel environment and its impact is shown in Fig. \ref{K_on_Eve}.

{\bf Lemma 1:} For a noisy channel, given that $d_{\rm x}$ and $d_{\rm y}$ are fractions of $\lambda,$ such that $d_{\rm x}=\frac{\lambda}{z_{\rm x}}$ and $d_{\rm y}=\frac{\lambda}{z_{\rm y}},$ then it holds that $K_{\rm x}\leq {z_{\rm x}}$ and $K_{\rm y}\leq {z_{\rm y}}$
%$A\le \lambda^2$ where $A$ is the area of the IRS.

\textit{Proof:} %The beamwidth of the reflected signal is inversely proportional to the IRS plate width and proportional to the wavelength, $\lambda$ \cite{Emil_2020}. 
It is known that for IRS plate width larger than $\lambda,$ the required local phase is coarsely quantized and will cause a mismatch between the desired reflection angle and the IRS array response in its far field \cite{Emil_2020}. It is apparent, then, to constrain the width of the entire IRS system within the bound of $\lambda$ such that $d_{\rm x}K_{\rm x}\leq \lambda$ and $d_{\rm y}K_{\rm y}\leq \lambda$ in order to minimize reflection mismatch. Simplifying the relations, completes the proof of Lemma 1. \hfill $\blacksquare$

%$A=LW\le \lambda^2$ $L=d_{\rm x}(K_{\rm x}-1)\le \lambda$ and $W=d_{\rm y}(K_{\rm y}-1)\le \lambda.$ By substituting $d_{\rm x}=\frac{\lambda{\rm c}}{z_{\rm x}}$ and $d_{\rm y}=\frac{\lambda{\rm c}}{z_{\rm y}},$ we have that $K_{\rm x}-1\le z_{\rm x}$
Provided that Lemma 1 is sustained, it is easy to see that $K\leq {z_{\rm x}}{z_{\rm y}}$ and the area of the entire IRS plate is upper bound by $\lambda^2.$

Due to the inverse relation between the beamwidth and the IRS plate width as given in \cite{Emil_2020}, we know that provided the bounds of Lemma 1 are sustained, the beamwidth reflected from the IRS will be smaller for increasing number of IRS elements, $K.$ This implies that the average secrecy rate of the system will increase for large values of $K$ since the reflected beam will be focused on Bob increasing its signal quality with less scatter for Eve. This invariably increases the average secrecy rate as shown in Fig. \ref{K_on_Eve}. However, both figures also shows that when Lemma 1 is not satisfied, the increase in the number of IRS elements is not guaranteed to improve average secrecy rate performance. This is because the reflected beamwidth is larger there is a mismatch between the reflection angle and the IRS array response thereby further empowering the signal received by Eve. It is important to note that in Fig. \ref{K_on_Eve1}, the channel quality is weak, %the location of Eve is close to Bob causing their channels to be highly correlated. 
nevertheless, the average secrecy rate tends towards the bound defined in Proposition 1. For Fig. \ref{K_on_Eve2}, the channel quality is strong %the channel of Eve and Bob are highly uncorrelated 
causing the significant difference between the bound in Proposition 1 and the actual secrecy rate. 

%The beamwidth becomes very small if the IRS area is large relative to $\lambda$.
%%%%%%%%%%%%%%%%%%%%%%%%%%%%%%%%%%%%%%%%%%

By examining the impact of the radius of the area where the transmit sensors are located, it is clear from Fig. \ref{radius} that increasing the radius improves on the average secrecy rate. This is primarily because increase in the radius allows the sensors to be scattered over a larger area ensuring that the $\boldsymbol{\Phi}_{\rm G}$ is not rank 1 and introducing greater variability to Eve. We note that since the design of the reflection coefficients, $\boldsymbol{\Theta},$ is focused on maximizing the quality of signal received at Bob, the impact of the variations obtained given that the rank of $\boldsymbol{\Phi}_{\rm G}$ is not 1 at Bob is reduced by the design of $\boldsymbol{\Theta}.$ 

\section{Conclusion} \label{sec4}
We have demonstrated the effectiveness of deploying a UAV-carried IRS for collecting sensor data from blackout noisy spaces while guaranteeing communications security. We have derived the optimal location as well as the reflection coefficients of the IRS elements to improve the secure data transmission performance. The optimal IRS location was determined through joint optimization of the trajectory of the UAV and the transmit beamforming. We have shown that the UAV follows a trajectory that can aid in sustaining the secrecy performance while the IRS can further assist secure communication by increasing phase disparity at the eavesdropper. The proposed algorithm is non-iterative which saves huge computation tasks compared to existing iterative procedures. Extensive simulation results demonstrate the effectiveness of the proposed approach.

\appendices
\section{}\label{Appendix_A}
%\section{Appendix 1: Proof of Proposition 1} \label{Appendix_A}
\comment{
The SNR at the main receiver (Bob) and Eve can be expressed as $$\gamma_{\rm B}=\frac{1}{\sigma_{\rm B}}|{\bf h}_{\rm B}^{\rm H}\bar{\boldsymbol{\Theta}}{\bf Gw}|^2=\frac{\bar{P}(\rho_0\varsigma)^2K^2}{d_{\rm RB}^2d_{\rm AR}^2},$$ 
$$\gamma_{\rm E}=\frac{1}{\sigma_{\rm E}}|{\bf h}_{\rm E}^{\rm H}\bar{\boldsymbol{\Theta}}{\bf Gw}|^2=\frac{\bar{P}(\rho_0\varsigma)^2|\zeta|^2}{d_{\rm RE}^2d_{\rm AR}^2}$$
where $\bar{P}=\frac{{\bf w}^{\rm H}{\bf w}}{\sigma_i}$,
\begin{multline*}
    \zeta=\frac{\sin(\frac{1}{2}K_{\rm x}A_{\rm x})\sin(\frac{1}{2}K_{\rm y}A_{\rm y})}{\sin(\frac{1}{2}A_{\rm x})\sin(\frac{1}{2}A_{\rm y})}\\
    \times \exp{\left(j\left(\theta_{\rm com}+\frac{A_{\rm x}(K_{\rm x}-1)}{2}+\frac{A_{\rm y}(K_{\rm y}-1)}{2}\right)\right)}.
    %\sum_{k_{\rm x}=1}^{K_{\rm x}}\sum_{k_{\rm y}=1}^{K_{\rm y}}\exp{(j[\theta_{\rm com}+(k_{\rm x}-1)\bar{d}_{\rm x}(\hat{a}_{\rm RB}^{\rm x}-\hat{a}_{\rm RE}^{\rm x})\\
       % +(k_{\rm y}-1)\bar{d}_{\rm y}(\hat{a}_{\rm RB}^{\rm y}-\hat{a}_{\rm RE}^{\rm y})])}
\end{multline*}
}
To provide a solution to Problem P5 given in \eqref{P5}, we express the lower bound for the distances using the reverse triangular inequality and variable change as
\begin{multline*}
    d_{\rm RE}^2d_{\rm AR}^2=\|\boldsymbol{\Omega}_{\rm E}-{\bf q}_{\rm R}\|^2\|{\bf q}_{\rm R}-\boldsymbol{\Omega}_{\rm A}\|^2\\
    \geq(\|\boldsymbol{\Omega}_{\rm E}\|-\|{\bf q}_{\rm R}\|)^2(\|{\bf q}_{\rm R}\|-\|\boldsymbol{\Omega}_{\rm A}\|)^2\\
    =\left(\underbrace{\frac{\|\boldsymbol{\Omega}_{\rm E}\|}{\|\boldsymbol{\Omega}_{\rm A}\|}}_{\bar{\Omega}_{\rm E}}-\underbrace{\frac{\|{\bf q}_{\rm R}\|}{\|\boldsymbol{\Omega}_{\rm A}\|}}_{\varepsilon}\right)^2\left(\underbrace{\frac{\|{\bf q}_{\rm R}\|}{\|\boldsymbol{\Omega}_{\rm A}\|}}_{\varepsilon}-1\right)^2\|\boldsymbol{\Omega}_{\rm A}\|^4\\
    =(\bar{\Omega}_{\rm E}-\varepsilon)^2(\varepsilon-1)^2\|\boldsymbol{\Omega}_{\rm A}\|^4 %~{\rm using ~variable~ change}
\end{multline*}
Similarly,
\begin{subequations}
\begin{align*}
    d_{\rm RB}^2d_{\rm AR}^2\geq (\bar{\Omega}_{\rm B}-\varepsilon)^2(\varepsilon-1)^2\|\boldsymbol{\Omega}_{\rm A}\|^4,\\
    \|{\bf q}_{\rm R}[n]-{\bf q}_{\rm R}[n-1]\|^2\geq (\varepsilon-\bar{\Omega}_q)^2\|\boldsymbol{\Omega}_{\rm A}\|^2;\\
    {\rm where}~\bar{\Omega}_q=\frac{\|{\bf q}_{\rm R}[n-1]\|}{\|\boldsymbol{\Omega}_{\rm A}\|}.
\end{align*}
\end{subequations}
Considering that the trajectory of the UAV is a sequential combination of its location at instantaneous $n$ samples, the objective of problem P4 can be scaled to obtaining the maximum value for each $n$ sample. The summation of these isolated optimal points provides optimal the objective value as defined in problem P5. Hence, by using variable change as defined above, problem P5 in \eqref{P5} can be rewritten as \eqref{P4b} $\forall{n\in[1,\dots,N]}$.
\begin{subequations}\label{P4b}
\begin{align}
    %\sum_{n=1}^N 
    \max_{\varepsilon}~& \log_2\left(\frac{1}{\beta}+\frac{\bar{P}M\rho_0^2\varsigma_{\rm B}^2K^2}{\beta(\bar{\Omega}_{\rm B}-\varepsilon)^2(\varepsilon-1)^2\|\boldsymbol{\Omega}_{\rm A}\|^4}\right) \label{C1},\\
    {\rm s.t.}~& \frac{1-\beta}{|\zeta|^2}+\frac{\bar{P}M\rho_0^2\varsigma_{\rm E}^2}{\beta(\bar{\Omega}_{\rm E}-\varepsilon)^2(\varepsilon-1)^2\|\boldsymbol{\Omega}_{\rm A}\|^4}\leq 0 \label{C2},\\
    ~& (\varepsilon-\bar{\Omega}_q)^2\|\boldsymbol{\Omega}_{\rm A}\|^2 \leq (Z\alpha)^2.
\end{align}
\end{subequations}
Problem \eqref{P4b} is differentiable and possibly non-convex due to \eqref{C1} and \eqref{C2}. However, let $\varepsilon^*$ and $(\lambda_1^*,\lambda_2^* )$ represent the primal and dual optimal variables with zero duality gap, the KKT conditions given in \eqref{KKT} must be satisfied.
\begin{subequations}\label{KKT}
\begin{align}
    \nabla f_0(\varepsilon^*)+\lambda_1^*\nabla f_1(\varepsilon^*)+\lambda_2^*\nabla f_2(\varepsilon^*) & = 0,\label{KKTa}\\
    \lambda_1^*f_1(\varepsilon^*) &= 0,\label{KKTb}\\
    \lambda_2^*f_2(\varepsilon^*) &= 0.\label{KKTc}
\end{align}
\end{subequations}
By using the functions from \eqref{P4b} where $f_0$ is the objective function and $f_1$ and $f_2$ are the constraint functions corresponding to \eqref{KKTb} and \eqref{KKTc} respectively, we note that $\lambda_1^*=f(\varepsilon^*,\lambda_2^*) {\rm~by~solving~\eqref{KKTa},}$ $\lambda_2^*=f(\varepsilon^*) {\rm~by~solving~\eqref{KKTb}~and~substituting~}\lambda_1^*.~$
Therefore, by solving \eqref{KKTc}, we obtain the cubic function $\varepsilon^3-b\varepsilon^2+c\varepsilon+d=0$ with discriminant $\Delta=(bc)^2+18(bcd)-4c^3-4b^3d-27d^2$; ($\varepsilon,~b,~c,~d$ has been presented in \eqref{va}). It is easy to see that the discriminant is less than $0$ which implies that the solution to the cubic function comprise of 2 complex conjugate pairs roots and one real root. Since we are interested in the coordinates located in the real plane, the only relevant solution is the real root as shown in \eqref{va}. Having obtained $\varepsilon$, the location of the UAV at the $n$th sample can be deduced from $\varepsilon=\frac{\|{\bf q}_{\rm R}\|}{\|\boldsymbol{\Omega}_{\rm A}\|}$ which can invariably be modified by expansion to Proposition 1.

\bibliographystyle{IEEEtran}
\footnotesize{

\bibliography{IEEEabrv, ref_AIRS}}%
%\IEEEtriggeratref{ref_no}
\end{document}